Title: Vibronic potentials in chemical physics: adiabatic approximation vs. supersymmetry
Author: Mladen Georgiev (Institute of Solid State Physics, Bulgarian Academy of Sciences
  1784 Sofia, Bulgaria)
Comments: 11 pages incorporating wording and 4(+1) figures, all pdf format
Subj-class: physics

We analyze the supersymmetric features of isolated double-well potentials, both symmetric ones and ones under an asymmetric perturbation. Our studies are in concert with results obtained earlier. Further on, a particular interest is paid to double and single valley potentials occurring in pairs on applying the adiabatic approximation to coupled electron – vibrational mode systems. Among the the latter are Holstein's small polarons and the vibronic (off-center) band species performing a finite-orbital-momentum helical motion. Our results seem to revive an earlier assertion for the existence of artifacts in the adiabatic approximation.

1. Introduction

Double wells, either symmetric or asymmetric appear on solving for electron-tunneling problems and/or classical transitions from one phonon-coupled electronic state to another state, in particular those pertaining to reaction rates [1]. Although certain approximations have been applied to obtaining energy eigenvalues, such as the quasiclassical (QCA) and adiabatic (AA) approximations, no exact solutions have been available. Nevertheless, the double-well method has been found fruitful for calculating the rate of a chemical reaction [2]. Some time ago, Keung, Kovacs and Sukhatme [3] obtained convincing results which dealt with symmetric double wells followed by a subsequent work by Gangopadhyaya, Panigrahi and Sukhatme et al. [4] on the asymmetric double wells, both from the viewpoint of susy quantum mechanics. There are several conceivable types, such as bottom-depth asymmetry (bda) and bottom-curvature asymmetry (bca). It is a matter of common belief that susy provides a powerful circumventing method for solving Schrödinger's equation in cases where the direct approach is less productive [5]. The calculated eigenvalue differences (tunneling splittings, zero-point reaction heats, etc.) between the component potentials of an asymmetric double well have been found comparable with those obtained by QCA, as well as with the numerical solution of Schrödinger's double-well equation (two site problem).

It is to be pointed out that while symmetric double wells in two-site problems apply to processes such as impurity dipole reorientation or polaron motion, asymmetric double valleys account for the rate of chemical reactions [1].

Nevertheless, no comparison has been made with AA which is often used in symmetric and asymmetric situations as the main approach most often involving first or second-order perturbation *a la* Rayleigh-Schrödinger. In turn, this paper is aimed at providing the missing information in order to see how it compares with the susy method which involves perturbative approaches too, albeit based on a different (logarithmic) method [4].

2. Susy approach to isolated double wells

To begin with, we outline the basic premises of susy quantum mechanics (susy-qm) which may not be outright familiar for most of the readers. Susy-qm is intended to describing superpartners and in order to obtain solutions for one of the pair components one has to take an operator square root of Schrödinger's equation of a pair. This is being done in the following way using operator techniques: If $V_-(x)$ is the relevant potential of the problem, then susy allows one to construct a "partner potential" $V_+(x)$ with eigenvalues in 1-to-1 correspondence with the excited-state eigenvalues of $V_-(x)$, that is, $E^-_n = E^+_{n-1}$ for any integer n. It does not include n=0, the $V_-(x)$ ground state, which has no paired correspondent on $V_+(x)$. The situation is typical for an asymmetric double well where the right-hand deeper component has one bound state below all the bound states on the left-hand shallower component.

It is customary to describe $V_-(x)$ in terms of its ground-state wave function. We assume that the wave function is known and that its corresponding eigenvalue is vanishing ($\hbar=2m=1$):

$$H_- \psi_0(x) \equiv [-d^2/dx^2 + V_-(x)] \psi_0(x) = 0 \qquad (1)$$

In terms of the ground state $H_-$ is also represented as:

$$H_- = -d^2/dx^2 + (\psi_0''/\psi_0) \qquad (2)$$

We introduce two operators: $A = d/dx - (\psi_0'/\psi_0)$, $A^\dagger = - d/dx - (\psi_0'/\psi_0)$ to obtain $H_- \equiv A^\dagger A$. Another semidefinite Hermitian operator is defined as

$$H_+ \equiv AA^\dagger = - d^2/dx^2 + V_+(x) \qquad (3)$$

$$V_+(x) = V_-(x) - 2 \, d/dx \, (\psi_0'/\psi_0) = - V_-(x) + 2 (\psi_0'/\psi_0)^2 \qquad (4)$$

At this stage, we introduce the superpotential

$$W(x) = - (\psi_0'/\psi_0) \qquad (5)$$

by means of which A and $A^\dagger$ are given by

$$A = d/dx + W(x), \quad A^\dagger = - d/dx + W(x) \qquad (6)$$

while

$$V_\pm(x) = W(x)^2 \pm W'(x) \qquad (7)$$

and the commutator $[A, A^\dagger] = 2W'(x)$.

Showing explicitly the $E^-_n = E^+_{n-1}$ correspondence, $H_+(A\psi_n^{(-)}) = AA^\dagger (A\psi_n^{(-)}) = A(A^\dagger A \psi_n^{(-)}) = AH_-(\psi_n^{(-)}) = E^-_n (A\psi_n^{(-)})$.

We further reproduce what has been obtained [3] by the susy equations for the symmetric case using the ground-state trial wave function

$$\psi_0(x) = \exp(-[x+x_0]^2) + \exp(-[x-x_0]^2): \qquad (8)$$

$$W(x) = 2[x - x_0\tanh(2xx_0)] \qquad (9)$$

$$V\pm(x) = 4[x - x_0\tanh(2xx_0)]^2 \pm 2[1 - 2x_0^2\mathrm{sech}^2(2xx_0)] \qquad (10)$$

for the super- and partner- potentials, respectively. The potentials $V_-(x)$, $V_+(x)$, and $W(x)$ for symmetric wells as they occur in supersymmetry problems are shown in Figures 1(a) and (b).

Finally we reproduce the results on supersymmetry potentials of asymmetric wells ($a \neq 1$), as follows:

$$W(x) = \{-2a(x-x_0)\exp(-a[x-x_0]^2)+2(x+x_0)\exp(-[x+x_0]^2\}/\{\exp(-[x+x_0]^2) +\exp(-a[x-x_0]^2)\} \qquad (11)$$

$$V_\pm(x) = W(x)^2 \pm W'(x) \qquad (12)$$

Here $a$ is the asymmetry parameter $a = 1$ for symmetric wells, $a < 1$ for asymmetric ones. The superpotentials $W(x)$ for both symmetries ($a=1$ and $a=0.4$) are compared in Figure 2. The form of the trial double-well eigenstate for the bca supersymmetry problem is:

$$\psi_0(x) = \exp(-a[x-x_0]^2) + \exp(-[x+x_0]^2). \qquad (13)$$

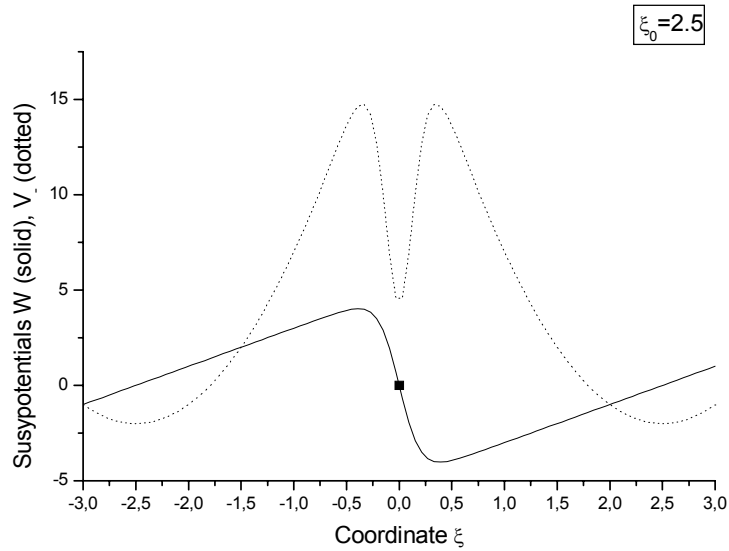

Figure 1(a): Partner potential $V_-(\xi)$ and superpotential $W(\xi)$ in susy quantum mechanics, as calculated for isolated symmetric double wells using the equations in Section 2. The superpotential extrema are seen to match the extrema on the partner potential very closely.

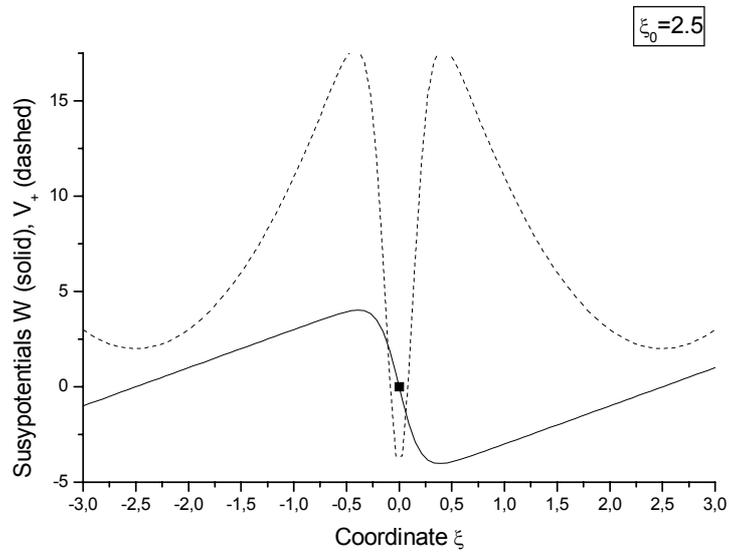

Figure 1(b): Partner potential $V_+(\xi)$ and superpotential $W(\xi)$ in susy quantum mechanics, as calculated for isolated symmetric double wells using the relevant equations of Section 2. The superpotential extrema match the extrema on the partner potential very well.

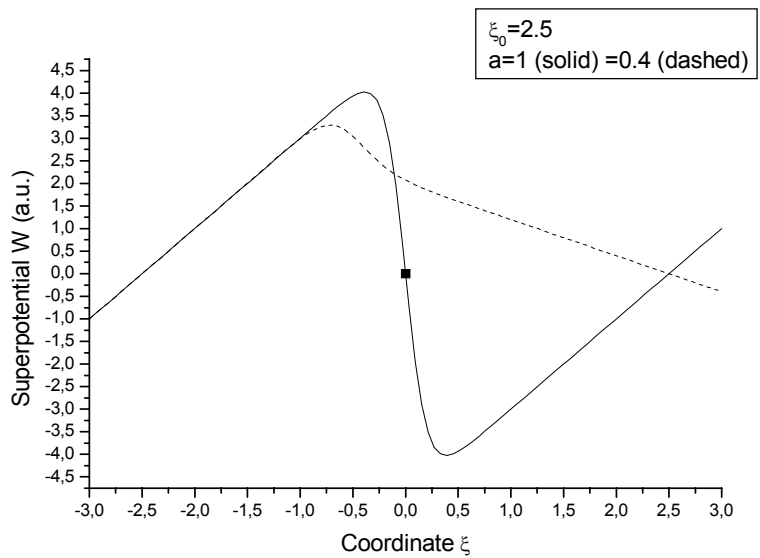

Figure 2: The superpotential $W(\xi)$ leading to an asymmetric double well (dashed) as compared with its symmetric counterpart (solid). The asymmetric shape seemingly tells a different story.

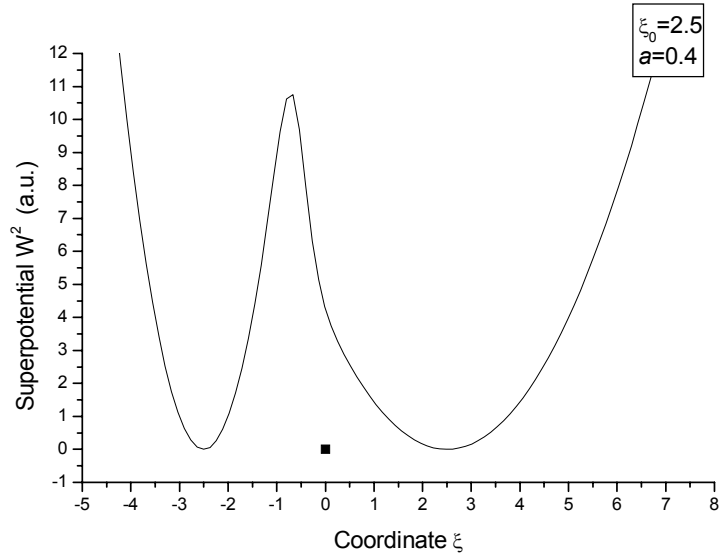

Figure 3: The squared superpotential of a bca asymmetric double well, calculated using equations (11) through (13), in particular the bca trial eigenstate (13).

### 3. Adiabatic approximation

#### 3.1. Symmetric wells pertaining to off-center polarons

We proceed to considering a specific example of phonon-coupled species to which most of our arguments apply rigorously: the off-center or vibronic polarons in solids.

Off-center displacements often appear in crystals as electronic charge carriers couple to centrosymmetric configurations which results in the breaking of their inversion symmetry. Pairs of lower-symmetry conformations appear, each pair producing two equal in magnitude and opposite in sign electric dipoles. These twin conformations are equivalent physically and tunneling transitions between them smear out the overall dipole thus restoring on the average the original inversion symmetry. Yet, the species are still polarizable in an electric field as is an s-wave atom. Consequently, nearby entities may be expected to pair via the analogue of the atomic Van der Waals forces in which the polarizability should incorporate the vibronic effect leading to the off-center instability. In this way off-center defects in crystals are expected to bind to each other to form dimers or larger aggregates – precursors to new phase formation. In the adiabatic approximation the above physics is described by double-well potentials which accounts for the overall importance ascribed in Section 2 to double-well processes.

Throughout this study we assume that the driving force of an off-center instability is the band pseudo-Jahn-Teller effect [6-8]. Accordingly we consider the following Hamiltonian:

$$H = \Sigma_{i,\beta} E_{i\beta}\, a_{i\beta}^\dagger a_{i\beta} + \Sigma_i \tfrac{1}{2} K_i q_i^2 + \Sigma_i G_i q_i (a_{i1}^\dagger a_{i2} + a_{i2}^\dagger a_{i1}) + T_N \qquad (14)$$

Here subscript "i" runs over the on-center sites hosting the vibronic polarons, while "β" labels two nearly degenerate opposite parity electron bands (β=1,2). $q_i$, $K_i$, and $G_i$ are the coordinates, spring and coupling constants, respectively, of the mixing (odd) vibrational mode, $a_{i\beta}^{\dagger}$, etc. are electron ladder operators. $T_N$ is the lattice kinetic energy operator.

To adiabatic approximation the electronic subproblem in (12) is dealt with by neglecting $T_N$ and regarding the nuclear coordinates $q_i$ as c-number parameters. After solving for the electronic subproblem, the obtained (adiabatic) electronic energy $E_L(q_i)$ is regarded as vibronic energy for dealing with the nuclear subproblem setting $H_{vib} = T_N + E_L(q_i)$. This program is to be implemented in greater detail below.

Using the adiabatic Hamiltonian $H_{AD} = H - T_N$ we solve Schrodinger's equation $H_{AD}\psi = E\psi$ by a superposition of one-particle electronic states in either electron band: $\psi = \Sigma_{i\beta} c_{i\beta} a_{i\beta}^{\dagger} |0\rangle$. A corresponding secular equation obtains for E whose roots are:

$$E_{U/L}(q_i) = \tfrac{1}{2} K_i q_i^2 \pm \tfrac{1}{2} \sqrt{[4G_i^2 q_i^2 + (E_{i2} - E_{i1})^2]} \tag{15}$$

To adiabatic approximation equation (15) describes a dual-branch vibronic potential energy of the electron-coupled oscillator at i-site. The lower branch $E_L(q)$ is a double-well potential with two symmetric minima at $\pm q_0$ given by (site index dropped) $q_0 = \sqrt{[(G/K)^2 - (E_{gu}/2G)^2]}$ where $E_{gu} = |E_2 - E_1|$ is the energy gap between the two narrow electronic bands (even and odd, mixed by the PJT active odd-parity vibration. Using (15) the vibronic Hamiltonian reads:

$$H_{vibL} = -(\hbar^2/2m)(d^2/dq^2) + E_L(q) \tag{16}$$

for the lower branch and a corresponding one for the upper branch:

$$H_{vibU} = -(\hbar^2/2m)(d^2/dq^2) + E_U(q) \tag{17}$$

The appearance of a dual-branch vibronic potential energy surface (15) is, perhaps, the most specific feature of the adiabatic approximation. The existence of the upper branch $E_U(q)$ is often disregarded less than correctly in studies of the double-well potential $E_L(q)$.

To first order we solve the pertinent Schrodinger equation by means of the linear combination

$$\chi(q) = A\chi_g(q-q_0) + B\chi_g(q+q_0) \propto A\exp(\xi-\xi_0)^2 + B\exp(\xi+\xi_0)^2 \tag{18}$$

where $\chi_g(q)$ is the ground-state wave function of a harmonic oscillator centered at $q = 0$:

$$\chi_g(q) = (K/\pi\hbar\omega)^{1/4} \exp(-[K/2\hbar\omega]q^2) \tag{19}$$

while $\hbar = h/2\pi$. Also $\xi = \sqrt{(K/\hbar\omega)}q$ is a dimensionless lattice coordinate to appear throughout. Inserting into $H_{vib}\chi(q) = E_{vib}\chi(q)$ two eigenvalues are obtained

$$E_{vib\ U/L} = (1 - S^2)^{-1} \{H_{RR} \pm (-H_{RL}) \pm S[H_{RR} \pm (-H_{RL})]\} \tag{20}$$

where

$$S = \exp(-\xi_0^2) \text{ with } \xi_0^2 = (K/\hbar\omega)q_0^2 = (2E_{JT}/\hbar\omega)(1 - (E_{gu}/4E_{JT})^2) \tag{21}$$

is the overlap integral between the two lateral harmonic-oscillator wave functions.

The remaining matrix elements in (20) are

$$H_{RR} = (\tfrac{1}{2} \hbar\omega)(1 + \xi_0^2 - I_1) \tag{22}$$

$$H_{RL} = (\tfrac{1}{2} \hbar\omega) S(1 - \xi_0^2 - I)$$

in which $I$ and $I_1$ are the following integrals:

$$I = \pi^{-1/2} \int_{-\infty}^{+\infty} \sqrt{[(8E_{JT}/\hbar\omega)\xi^2 + (E_{gu}/\hbar\omega)^2]} \exp(-\xi^2)\, d\xi \tag{23}$$

$$I_1 = \pi^{-1/2} \int_{-\infty}^{+\infty} \sqrt{[(8E_{JT}/\hbar\omega)(\xi+\xi_0)^2 + (E_{gu}/\hbar\omega)^2]} \exp(-\xi^2)\, d\xi$$

Using (22) the ground-state tunneling splitting is:

$$\Delta E = E_{vibU} - E_{vibL} = 2(1-S^2)^{-1}(SH_{RR} - H_{RL}) = [\sinh(\xi_0^2)]^{-1}(H_{RR} - S^{-1}H_{RL}) \tag{24}$$

By means of the steepest-descent method, which works at $4E_{JT} \gg E_{gu}$ (small polaron), we obtain $I \sim E_{gu}/\hbar\omega$, $I_1 \sim 4E_{JT}/\hbar\omega$ and since $\xi_0^2 \sim 2E_{JT}/\hbar\omega$, we finally get for small polarons

$$\Delta E = [2\sinh(\xi_0^2)]^{-1} E_{gu}(1 - E_{gu}/4E_{JT}) \sim E_{gu}\exp(-2E_{JT}/\hbar\omega) \tag{25}$$

At the other extreme (large polaron) $4E_{JT} \geq E_{gu}$, conventional wisdom for large polarons leads to the tunneling splitting being given by the undressed electron bandgap: $\Delta E \sim E_{gu}$. Here and above $E_{JT} = G^2/2K$ is the Jahn-Teller or coupling energy. For small polarons, $E_{JT}$ is about the height of the interwell barrier.

A comparison with familiar small polaron (Holstein) formulae is impending. Vibronic intraband mixing effects can also be considered based on a Hamiltonian similar to (14). The analysis leads to a tunneling splitting as above with $E_{gu}$ standing for the bandwidth. At large coupling strengths (or small bandwidths) equation (25) obtains where $\Delta E$ is the itinerant polaron bandwidth.

In this respect it is worth noting that while the susy analysis leads to partner potentials of the forms (10)-(13), the adiabatic approach to the dual-branch vibronic potential energy leads to very different forms (15). This is apparently due to the choice of a ground-state trial wave function, even though the ground state of (14) is described by a wave function (18) similar to (10). This comes to show that more has to be done to elucidate the ground-state properties of the system described by equations (15).

### 3.2. Solving for the susy problem of vibronic potential energy

Given the doubts in the consistency of the trial wave function (13) to solving for the adiabatic problem, then one should regard (7) as a differential equation for the superpotential $W(\xi)$:

$$W'(\xi) = W(\xi)^2 - E_{U/L}(\xi) \tag{26}$$

This is a Riccati equation [11] which should give the superpotential $W(\xi)$ and then from

$$V_+(\xi) = W(\xi)^2 + W'(\xi) \tag{27}$$

the partner potential $V_+(\xi)$ too. We see the partner potentials $V_\pm(\xi)$ to be closely interrelated.

The differential equation (26) can be converted into a second-order linear equation by means of the substitution

$$U(\xi) = \exp(-\int W(\xi)d\xi) \qquad (28)$$

yielding a vibronic Schrödinger equation with zero eigenvalue

$$U(\xi)'' - E_{U/L}(\xi)U(\xi) = 0 \qquad (29)$$

Now from $U(\xi) \propto \exp(-[x-x_0]^2)$ and equation (29) we get $W(\xi) \propto 2[x - x_0]$ which underlines the elastic force character of the superpotential in the particular problem.

The adiabatic potentials $E_{U/L}(\xi)$ are each composed of anharmonic parabolas at renormalized bottom curvatures preserving the oscillator mass. $E_U(\xi)$ is a normal parabola centered at $\xi_0 = 0$ at renormalized frequency $\omega_{Uren} = \omega_{bare}(1 + 4E_{JT}/E_{gu})$. In contrast, $E_L(\xi)$ is a double well parabola with a central barrier at $\xi = 0$ and lateral valleys bottoming at $\pm\xi_0$ (cf. (21) where $\xi_0 = \sqrt{\{(2E_{JT}/\hbar\omega)[1 - (E_{gu}/4E_{JT})^2]\}}$ at $(E_{gu}/4E_{JT}) \ll 1$ (small-polaron condition) with renormalized bottom frequencies $\omega_{Lren} = \omega_{bare}\sqrt{[1 - (E_{gu}/4E_{JT})^2]}$. This gives two trial harmonic oscillator wave functions: $\chi_{Lgren}(\xi) = \chi_{Lgren}(\xi-\xi_0) + \chi_{Lgren}(\xi+\xi_0)$ for $E_L(\xi)$ and $\chi_{Ugren}(\xi)$ for $E_U$. Although the above representation of $E_{Lren}(\xi)$ and $E_{Uren}(\xi)$ is only approximate, it will be good for processes not too high above the barrier top. For example, it will hold good for the ground states of the adiabatic vibronic potentials. See Figure 4 for graphic representations.

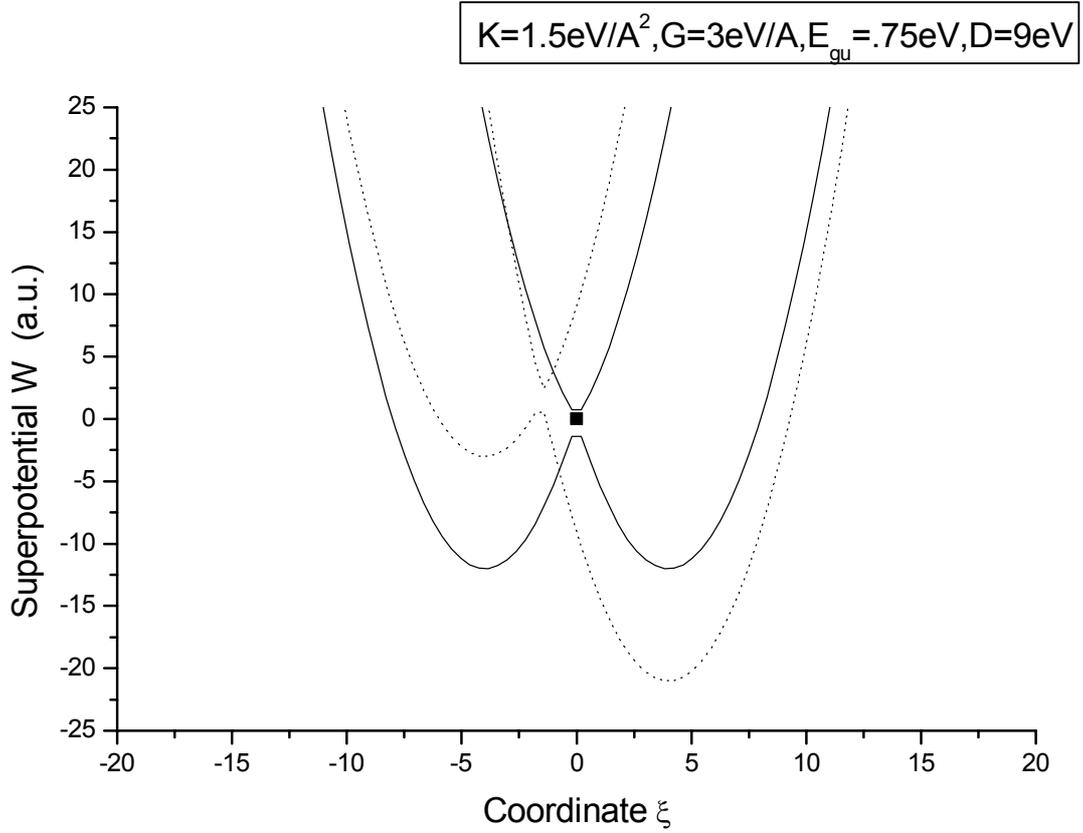

Figure 4: The symmetric (solid) and bda asymmetric (dashed) adiabatic potentials, as calculated using the equations of Section 3. Note that the lower and upper branches always appear in pairs, unlike isolated double wells appearing in previous Figures. In particular, the double wells on lower branches are formed by virtue of the electron-vibrational mode coupling which creates two (a)symmetric valleys. The asymmetry arises as a mode-independent dipolar mixing energy is added to the interaction Hamiltonian as explained in Section 3.3.

### 3.3. Asymmetric wells

We consider a few Hamiltonians in which electrostatic monopoles couple to a quantized field of off-center polarons. These are based on the original symmetric Hamiltonian (14) which may be made asymmetric by introducing a coupling term or different curvatures for the left- and right- hand valleys. In another example the vibronic tunneling splitting is widened. However, this type conserves the original symmetry of Hamiltonian. Another one incorporates an electrostatic dipole coupling term $D = -2\mathbf{p}_{\alpha\beta} \cdot \mathbf{F}$ in (14) so as to introduce different depths for the double-well potential valleys:

$$H_M = \Sigma_{i\beta} E_\beta a_\beta^\dagger a_{i\beta} + \Sigma_i \tfrac{1}{2} Kq^2 \pm \Sigma (Gq + \tfrac{1}{2}D)(a_1^\dagger a_2 + a_2^\dagger a_1) + T_N \qquad (30)$$

Its bda adiabatic potential is

$$E_{U/L}(q) = \tfrac{1}{2} Kq^2 \pm \tfrac{1}{2} \sqrt{[(2Gq + D)^2 + (E_{gu})^2]} \qquad (31)$$

Equations (15) and (31) represent examples of symmetric and asymmetric double wells of the vibronic problem. The vibronic potentials should be analyzed by adiabatic approximation and susy qm.

## 4. Comments

Considering an asymmetric double well composed of two valleys separated by an interwell barrier, as in Figure 3, we stress that in the absence of an interwell coupling interaction through tunneling there is a 1-1-correspondence between the levels on the individual wells, except for the ground-state level on the deeper component which remains unpaired. The difference between the ground state levels on the deeper and shallower components is regarded as reaction heat Q at 0 K. As the tunneling coupling is "switched on", the individual levels split into pairs of tunneling components which are common for the individual valleys. Again, the deep level on the asymmetric well is unpaired since there are no tunneling transitions to the shallow component starting from that deep level. For this reason, the deep level does not change to first order unlike all the remaining levels and it is easy to predict that the position of the deep level will be one phonon quantum energy below the first excited level on the deeper component: $Q = n\hbar\omega$, usually for $n = 1$. Under these conditions, the deep level will be unpaired in a susy pattern, while a 1-1- correspondence will be established between the tunneling levels. Now, the difference between the ground state split levels on the two wells will have very little to do with the tunneling splitting, in either symmetric and asymmetric double wells which is often mistaken in literature. The correct definition for a ground state tunneling splitting will be the difference between the first excited level and the ground state level in a symmetric double well, or between the second excited level and the first excited level above the deeper level on the asymmetric double well.

Comparing the present Figures 1 (a) and (b) with the graphic material on display in Reference [3], we see a general correspondence except for the abrupt central dip splitting the interwell peak in Figure 1 (a). This makes the potentials $V_-(\xi)$ and $V_+(\xi)$ similar though not akin to each other. The central dip feature in Fihure 1 (a) though missing in Reference [3] is remindful of earlier analytic work pointing to artifacts in the adiabatic approximation [12]. Indeed, through carefully defining and removing the electronic nonadiabaticity operator, the author has observed a central surge on top of the interwell maximum in the double well potential of a phonon coupled two level system. Our observation is of a surge-like dip on top of that same potential. One is tempting to check whether the two opposite-sign surges do not compensate each other under adiabatic conditions.

Finally, we conclude that there is more interesting work ahead to elucidate the super-symmetric and adiabatic features of vibronic potentials in chemical physics.